\documentclass[letterpaper, 10 pt, conference]{IEEEtran}  

\IEEEoverridecommandlockouts                              

\usepackage{balance}
\usepackage{graphicx}
\usepackage{url}
\usepackage{tabularx}
\usepackage{makecell}

\usepackage{siunitx}
\usepackage{subcaption}
\usepackage{booktabs}
\usepackage{hyperref}
\usepackage{multirow}
\usepackage{float}
\usepackage{comment}
\usepackage{layout}
\usepackage{geometry}
\title{\vspace{0.6cm} \LARGE \bf How is the Pilot Doing: VTOL Pilot Workload Estimation by Multimodal Machine Learning on Psycho-physiological Signals\vspace{-0.4cm}}

\author{%
  \IEEEauthorblockN{%
    \parbox{\linewidth}{\centering
      Jong Hoon Park\IEEEauthorrefmark{1}\IEEEauthorrefmark{2},
      Lawrence Chen\IEEEauthorrefmark{1}\IEEEauthorrefmark{2},
      Ian Higgins\IEEEauthorrefmark{2},
      Zhaobo Zheng\IEEEauthorrefmark{3},
      Shashank Mehrotra\IEEEauthorrefmark{3},
      Kevin Salubre\IEEEauthorrefmark{3},
      Mohammadreza Mousaei\IEEEauthorrefmark{2},
      Steven Willits\IEEEauthorrefmark{2},
      Blaine Levedahl\IEEEauthorrefmark{3},
      Timothy Buker\IEEEauthorrefmark{3},
      Eliot Xing\IEEEauthorrefmark{2},
      Teruhisa Misu\IEEEauthorrefmark{3},
      Sebastian Scherer\IEEEauthorrefmark{2},
      and Jean Oh\IEEEauthorrefmark{2}%
    }%
  }%
  \IEEEauthorblockA{%
    \IEEEauthorrefmark{1}These authors made equal contributions to this work\\
    \IEEEauthorrefmark{2}Robotics Institute, Carnegie Mellon University\\
    \IEEEauthorrefmark{3}Honda Research Institute USA, Inc%
  }%
}

\begin{document}
\vspace{10pt}
\maketitle
\thispagestyle{empty}
\pagestyle{empty}

\begin{abstract}
Vertical take-off and landing (VTOL) aircraft do not require a prolonged runway, thus allowing them to land almost anywhere. In recent years, their flexibility has made them popular in development, research, and operation. When compared to traditional fixed-wing aircraft and rotorcraft, VTOLs bring unique challenges as they combine many maneuvers from both types of aircraft. Pilot workload is a critical factor for safe and efficient operation of VTOLs. In this work, we conduct a user study to collect multimodal data from 28 pilots while they perform a variety of VTOL flight tasks. We analyze and interpolate behavioral patterns related to their performance and perceived workload. Finally, we build machine learning models to estimate their workload from the collected data. Our results are promising, suggesting that quantitative and accurate VTOL pilot workload monitoring is viable. Such assistive tools would help the research field understand VTOL operations and serve as a stepping stone for the industry to ensure VTOL safe operations and further remote operations.
\end{abstract}

\section{INTRODUCTION}
The interest in urban air mobility (UAM), such as VTOL aircraft, has steadily increased through the past few decades. Combining the advantages of a rotorcraft and fixed-wing aircraft, VTOLs can take off and land from the same place without the need for a long runway \cite{jagdale2018vertical}. VTOLs are advantageous in urban and short-distance flights as they allow faster cruise speeds, better transport capabilities, and ease of landing and take-off \cite{silva2018vtol}. The more modern electric or hybrid VTOLs provide additional benefits, including lower emissions and noise levels, which are highly beneficial for UAM \cite{xiang2023autonomous}. With these advantages, VTOLs introduce new modes of mobility, expanding the ways individuals can travel.

However, despite their advantages, VTOLs present different challenges and considerations compared to existing aviation modes. For example, due to their better ability to take off and land for short-distance flights, most current commercial VTOL designs are smaller in size compared to their fixed-wing counterparts. Thus, most commercial VTOLs are designed for a single pilot on board. This reduces redundancy in the cockpit when pilot workload reaches critical levels during a flight \cite{faulhaber2019crewed}. Moreover, VTOLs combine challenging maneuvers from both rotorcraft and fixed-wing aircraft, such as both vertical and rolling landings. Many potential VTOL pilots come from either rotorcraft or fixed-wing backgrounds, and frequent transitions between the maneuver habits may add a significant workload \cite{zaludin2019automatic}. VTOLs also bring unique challenges, such as vertical and horizontal flight transitions. These novel tasks require additional knowledge about flight dynamics and control, and the related workload is critical but poorly understood.


In this work, we demonstrate pilot workload estimation for VTOL operation through psycho-physiological sensing. We design flight tasks, taking into consideration the uniqueness of VTOL maneuvers, and collect data and associated workloads. We also integrate a complete set of sensors that provides design guideline for future VTOL pilot workload estimation. To the best of our knowledge, this work is the first in the field to investigate psycho-physiological measurements of workload for pilots of VTOL aircraft. We also contribute to the understanding of the predictive power of almost all available wearable sensor signals in such tasks.

The rest of the paper is organized as follows. In Section~\ref{sec:related-work}, we discuss some preexisting literature on workload estimation. In Section~\ref{sec:user-study}, we introduce details about the user study, including the simulation platform, sensor setup, and flight task design. In Section~\ref{sec:signal-processing}, we go over the various signal processing and feature extraction methods performed on the raw data. Then, in Section~\ref{sec:statistical-analysis}, we show some preliminary results of the statistical analysis. In Section~\ref{sec:workload}, we describe the machine learning algorithms used to estimate workload. Finally, we conclude the paper and discuss the limitations and future research directions in Section ~\ref{sec:conclusion}. 

\section{RELATED WORK} \label{sec:related-work}

Workload measurement and management for VTOL piloting is vital for safe and efficient commercial operation \cite{liu2016cognitive}. There has been some existing research on this topic for single-pilot fixed-wing aircraft. It was found that subjective measurements like NASA-TLX were able to effectively measure task demands \cite{battiste1988transport}. However, during normal operations, pilots are already highly occupied with their primary tasks and would not have the capacity for questions or workload self-reports, especially during or immediately after the task(s) being evaluated for workload. Continuous, non-invasive, and automatic measurements are needed for practical monitoring. Wilson analyzed pilots' mental workload during flight using psycho-physiological measures \cite{wilson2002analysis}. Ten pilots flew for about 90-minute scenarios while their heart rate, eye blinks, electrodermal activity, and electrical brain activity were recorded. Cardiac and electrodermal measures were found to be highly correlated with task demands, and blink rates decreased during the more highly visually demanding tasks. Eye gaze was more thoroughly studied later for commercial single-pilot operations \cite{faulhaber2019eye}. Participants flew short approach and landing scenarios with or without the support of a second pilot, and fixation increased while dwell duration decreased during single pilot operations. Machine learning tools are also used to identify pilot mental workload. Mohanavelu et al. used support vector machine (SVM), k-nearest neighbors (kNN), and linear discriminant analysis (LDA) to detect normal and high workload \cite{mohanavelu2022machine}. They found out that heart rate variance features were more predictive than electric brain activity features, and the workload was better predicted during the cruise and landing phase than the takeoff phase. Taheri et al. used EEG and similar machine learning algorithms to classify low, medium, and high workloads \cite{taheri2023using}. They used power spectral density and log energy entropy to select the features and achieved improved performance. Existing literature demonstrates the possibility of using psycho-physiological measurements to detect their cognitive workloads. However, there are two main limitations of existing literature. They all focus on fixed-wing operations. The workload from unique VTOL maneuvers, such as transitions between rotorcraft and fixed-wing modes, have not been researched. Most existing work also utilize only one type of signal, while some use a combination of two types of signals. A dataset with a more complete set of wearable sensors could provide insight into which combination of sensors should be integrated in the future cockpit. 

\section{USER STUDY}\label{sec:user-study}

\subsection{Simulation} 
\begin{figure}[!tbp]
  \begin{minipage}[b]{0.27\textwidth}
    \includegraphics[width=\textwidth]{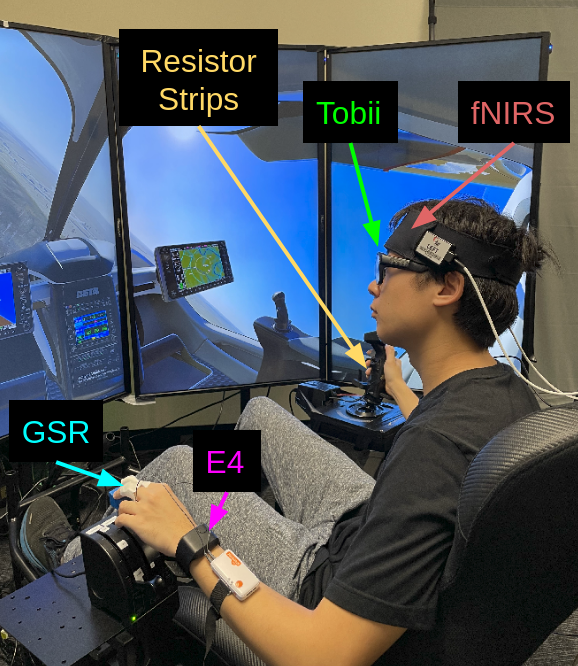}
    \caption{Flight Simulator}
    \label{fig:Simulator}
  \end{minipage}
  \hfill
  \begin{minipage}[b]{0.20\textwidth}
    \begin{subfigure}[b]{\textwidth}
      \includegraphics[width=\textwidth]{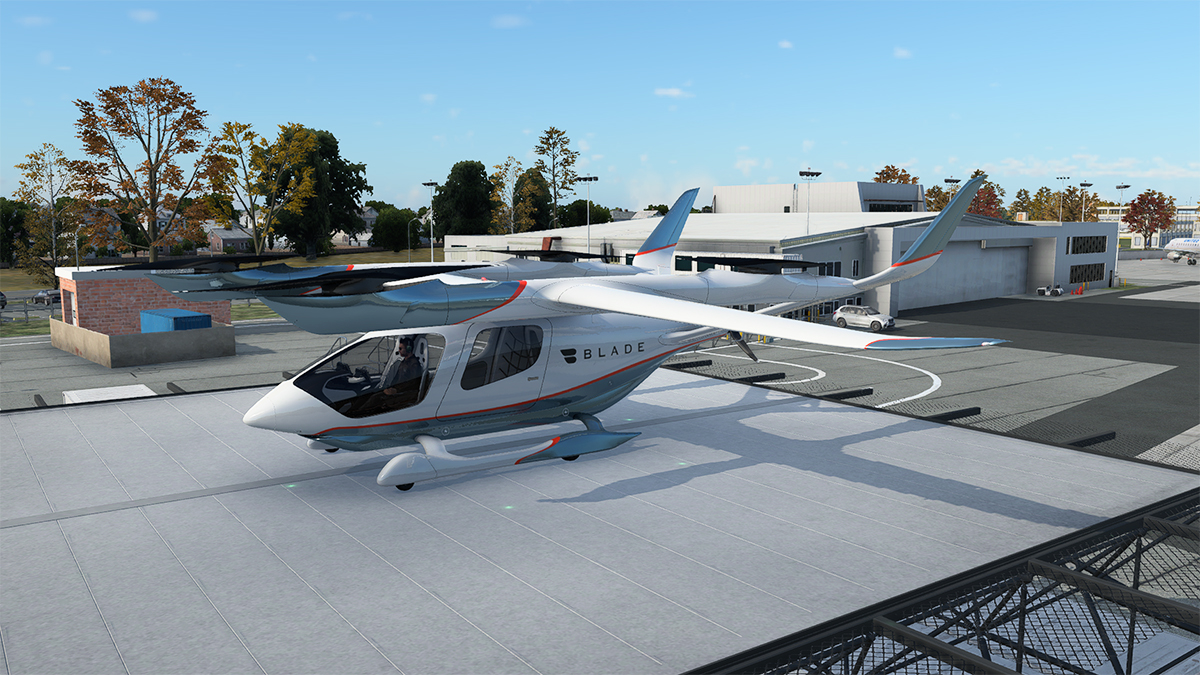}
      \caption{VTOL in X-Plane}
      \label{xplaneVTOL}
    \end{subfigure}
    \vspace{0.5em}
    \begin{subfigure}[b]{\textwidth}
      \includegraphics[width=\textwidth]{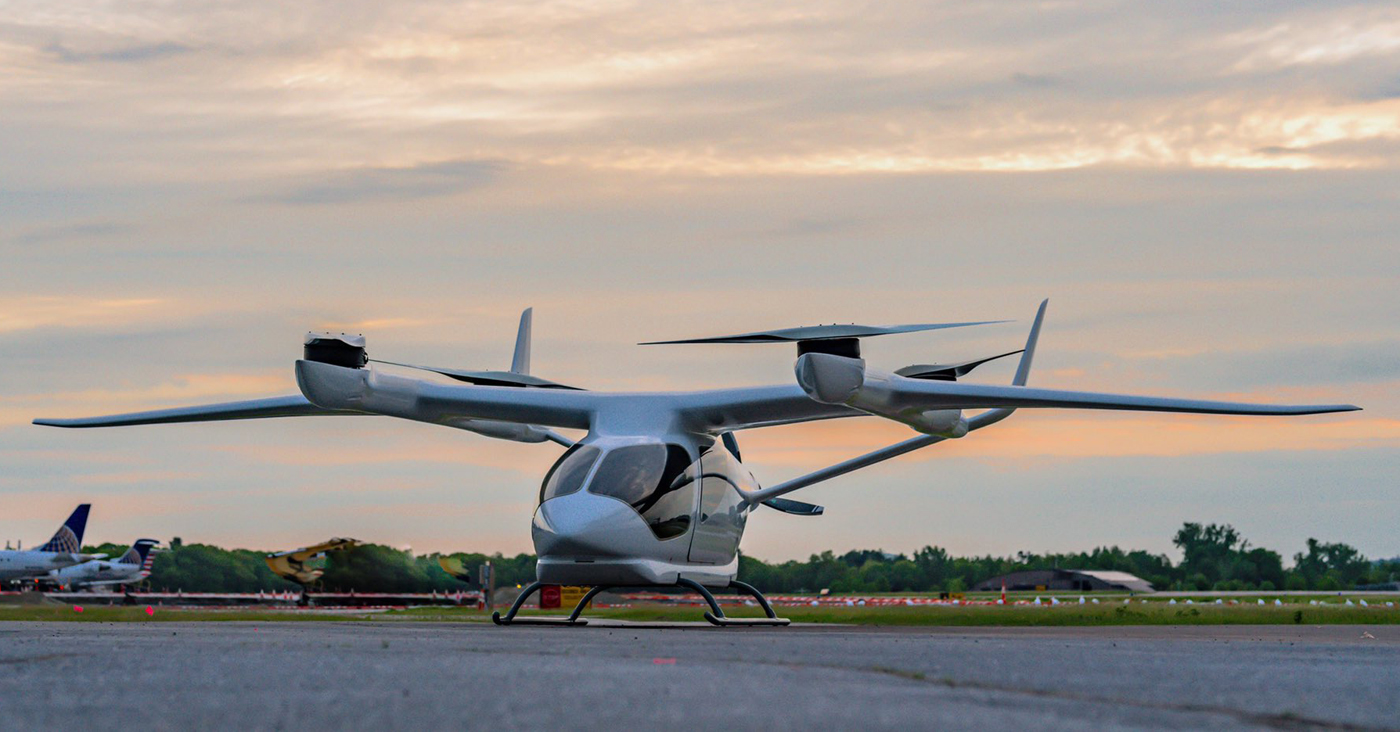}
      \caption{VTOL in real life}
      \label{realVTOL}
    \end{subfigure}
    \caption{VTOLs in simulator and real life}
    \label{VTOL rendering}
  \end{minipage}
\vspace{-0.7em}
\end{figure}

We integrate a medium-fidelity aviation simulation platform with X-Plane 12 \cite{xplane}. The flight scenes are rendered on 5 vertically oriented screens, resulting in a total horizontal field of view of 225 degrees, as shown in Fig. \ref{fig:Simulator}. The wide field of view provides an immersive flying experience and aims to give the pilot more situational awareness. For the aircraft model, we choose the Beta Technologies ALIA-250 electric VTOL \cite{BETA-Alia} as it has an off-the-shelf model in X-Plane 12. It has 4 vertical propellers and 1 horizontal propeller, as can be seen in Fig. \ref{VTOL rendering}. The controls consist of a throttle quadrant, a joystick, and a pair of rudder pedals. The throttle quadrant has two sticks responsible for the vertical and horizontal propellers. We customize the vertical stick to adaptively set the rate of ascent/descent of the aircraft. The joystick is used to control the roll and pitch of the aircraft, while the rudder pedals are responsible for the yaw. The pedals are also used to control the aircraft's differential brakes, allowing for turning and stopping on the ground. This pilot control configuration is not specific to Beta or other aircraft.

\subsection{Sensing Framework}
We integrate a multimodal sensing framework to record the pilot's psycho-physiological signals. The sensors consist of an Empatica E4 wristband, a Shimmer GSR+ device, a Microsoft Kinect V2, a BIOPAC fNIRS headband, a pair of force-sensitive resistor strips on the joystick, and a pair of Tobii eye tracking glasses. Many of these sensors have demonstrated prediction power in prior work \cite{alaimo2020aircraft,doyon2014effects,feng2018comprehensive,causse2019influences}, though force sensing and body pose tracking have been relatively unexplored.

The E4 sensor is a wristband we wrap around the participant's left wrist. It uses photoplethysmography (PPG) to measure blood volume pulse (BVP), heart rate (HR), and interbeat interval (IBI). In addition, it also measures body temperature and wrist acceleration.

The Shimmer GSR+ device has two electrodes which we attach to the index and middle fingers of the participant's left hand. It measures the galvanic skin response (GSR), which reflects sweat gland activity. It is worth noting that the E4 sensor measures GSR as well, but we use the Shimmer GSR+ for improved data quality and fewer motion deficits. Heart rate and skin conductance are the most common peripheral physiological signals and have been proven to strongly correlate with arousal and mental workload \cite{shimomura2008use}.

We collect brain activity data from the fNIRS optical brain imaging sensor. This sensor looks like a hairband and is worn on the forehead. It uses 4 infrared light emitters, 10 detectors, and 18 optodes to noninvasively measure oxygen levels in the prefrontal cortex. Compared to EEG, which is another popular sensor in brain activity measurement for cognitive state sensing, fNIRS is less invasive and more comfortable with comparable prediction power \cite{masters2020investigating}.

The Tobii Pro 3 eye tracking glasses have the shape of normal glasses with integrated IR-LED emitters and microcameras. It has a front-facing scene camera with a 106° field of view to capture the pilot's POV during the flight. During recording, each frame is annotated with the 2D coordinates of the pilot's gaze.

Tactile information on the controllers is strongly related to maneuvers, which is a strong indicator of physical workload \cite{smith2022decomposed}. We strap two force sensing resistors to the handle of the joystick to measure the amount of pressure applied on those areas. This setup aims to capture factors like grip strength.

The 3D spatial upper body poses of participants are collected using a Kinect V2 sensors. The upper body joints, including the shoulders, elbows, wrists, and head, are the primary moving parts when operating aircraft as hands control the joystick and throttles, while the head constantly scans the scene for points of interest.


We used XPlaneROS, an X-Plane wrapper application, that provides functionality for extracting aircraft data from the simulator and feeding control commands to control the aircraft \cite{baijal_patrikar_moon_scherer_oh_2021}. Using it, we collected flight control inputs made by the participants through the joystick and throttles, as well as spatio-temporal positions and orientations of the aircraft.

\subsection{Flight Tasks}


We design the flight tasks to represent core aspects of VTOL commercial operations. We use the private powered lift airman certification standards (ACS) by the Federal Aviation Administration (FAA) as a guideline while designing tasks, which is based on the risk management handbook by FAA \cite{RiskManagementHandbook}. The tasks include taxiing, vertical takeoff, transition from vertical flight to horizontal flight, maintaining altitude, climbing, turning to heading, landing in fixed-wing mode, and landing vertically. Three certified pilots (1 private, 1 commercial, and 1 airline) with extensive flight experiences provided ratings for the expected difficulty of the tasks to achieve well-distributed workload levels from the participants. They rated the expected task difficulty from 1 to 6, with 1 requiring only a guided mental process and 6 requiring high dynamic mental process plus dynamic physical action close to ground obstructions.
A complete list of the tasks and their expected difficulty is shown in Table \ref{tab:flighttasks}.

\begin{table*}[t]
\caption{Flight Tasks and Expected Difficulty}
\label{tab:flighttasks}
\centering
\begin{tabular}{|p{1.2cm}|p{9.0cm}|p{1.3cm}|p{1.1cm}|p{0.7cm}|p{0.7cm}|p{0.7cm}|}
\hline
\textbf{ACS Ref} & \centering \textbf{Task} & \textbf{Expected Difficulty} & \textbf{Survey Group} & \textbf{Task Group} & \textbf{Flight Group}\\
\hline
IV.A & Vertical takeoff to 1600 MSL. Hold position and heading 280 for 30 seconds. & 3 & 1 & 1 & A\\
\hline
IV.B & Maintain altitude, rotate to heading 360. Hold for 10 seconds. & 3 & 1 & 1 & A\\
\hline
V.A & Transition from hover to forward flight. Use full horizontal thrust until at 100 knots. Then turn off vertical thrust. Maintain heading and altitude. & 4 & 1 & 2 & A\\
\hline
V.C & Climb to 3000 feet at 80 knots using full horizontal thrust. Once at 3000 feet, accelerate to 100 knots. Maintain heading. & 3 & 1 & 3 & A\\
\hline
VI.B & Turn to 090 with 30 degree bank. Maintain altitude and 100 knots. & 3 & 1 & 3 & A\\
\hline
VI.B & Turn to 180 with 30 degree bank. Maintain altitude and 100 knots. & 3 & 1 & 3 & A\\
\hline
VI.C & Perform a steep turn to the right, using 45 degree bank, back to 180. Maintain altitude. & 4 & 2 & 4 & A\\
\hline
VI.C & Perform a steep turn to the left, using 45 degree bank, back to 180. Maintain altitude. & 4 & 2 & 4 & A\\
\hline
VI.C & Perform slow flight at 75 knots for 30 seconds. Maintain heading 180 at 3000 feet. & 4 & 2 & 4 & A\\
\hline
VI.B & Transition back to vertical flight. Pull back on horizontal thrust all the way and set vertical speed to 0. Maintain heading 180 at 3000 feet. & 5 & 2 & 5 & A\\
\hline
VI.B & Hold heading and position for 30 seconds. & 3 & 2 & 6 & A\\
\hline
V.C & Fly back to KAGC. Follow the blue heading bug. & 2 & 2 & 6 & A\\
\hline
V.A, V.D & Descend to 2300 feet. Enter left traffic pattern for runway 28. & 4 & 3 & 7 & A\\
\hline
V.D, V.E & Hover over runway 28 markers at 1600 feet. Land vertically full stop. & 5 & 3 & 7 & A\\
\hline
IV.A, IV.C & Takeoff vertically and hover at 1300 feet. Air taxi to front of tower. Then return to runway 28. Stay above the taxiway while air taxiing. & 6 & 4 & 8 & B\\
\hline
V.B & Hover on the centerline of runway 28, maintaining heading 280 and 1300 feet, for 30 seconds. Then land. & 6 & 4 & 8 & B\\
\hline
IV.A & Maintain altitude, rotate to heading 180. Hold for 10 seconds. & 3 & 4 & 9 & B\\
\hline
IV.B & Maintain altitude, rotate to heading 090. Hold for 10 seconds. & 3 & 4 & 9 & B\\
\hline
V.E, II.D & Land. Taxi on the ground to the front of tower. Then return to runway 28. & 6 & 4 & 10 & B\\
\hline
V.B & Perform a short field rolling takeoff, using vertical thrusters if necessary. Enter left traffic pattern and set up for approach on runway 28. & 5 & 5 & 11 & C\\
\hline
V.E & Perform rolling landing touch and go on runway 28. & 5 & 5 & 11 & C\\
\hline
V.H & Transition back to vertical flight. Land between two adjacent hangars at KAGC. & 6 & 5 & 12 & C\\
\hline
V.H & Takeoff in this confined area, air taxi to front of tower and land full stop. & 5 & 5 & 13 & C\\
\hline
\end{tabular}\vspace{-0.4cm}
\end{table*}

\subsection{Experimental Protocol} 
Each experiment takes approximately 2 hours from participant arriving to leaving, with the actual flight time taking roughly 1 hour. Two researchers oversee each participant, with one moderator facilitating the tasks and overall procedures and one technical operator overseeing the sensor modalities and data collection.

When each participant arrives, we first describe the purpose of the study and obtain their written consent to participate. Next, we show and describe each sensor. This is followed by a practice session in which the participant is given time to learn the flight simulator controls and conduct some practice maneuvers. During this session, we guide them through a vertical takeoff, a rolling takeoff, and transitions between vertical and fixed-wing modes. They are also given a few minutes to perform maneuvers on their own. Then, we attach the sensors and conduct a physiology baseline collection for roughly 30 seconds, during which participants are instructed to stare at a dot on an otherwise empty screen. Physiological signals such as heart rate, skin conductance, and grip force depend on physical characteristics like weight, height, and strength \cite{appelhans2006heart}, so we aim to use the baseline data to remove the individualized differences. 

After the baseline data collection, we start the experimental sessions with the flight tasks. To initiate each task, the moderator plays an audio recording with flight instructions while the technical operator records the start timestamp. If the participant fails a certain task, for example by crashing, this event would be recorded, and the simulator would be set to a state suitable for participants to continue the following flight task. All footage of the experiment is manually reviewed and annotated so that each task has an accurate start/end time, as well as a label indicating whether or not it was completed successfully.

Throughout the experiment, participants may gain better skills and familiarity with the system, which may ease their workload \cite{boehm2007pilot}. At the same time, fatigue may increase with flight time and affect workload \cite{borghini2014measuring}. To avoid task order biases, we counter-balance the order of the flight tasks. However, continuous flight is still needed for the pilots to have a natural and immersive experience to reflect the appropriate workload. Therefore, we group the tasks into three flights, each a continuous sequence that starts and ends at the runway. We assign each participant one of the 6 permutations of the 3 flights, so that the bias of exposure is reduced while the simulation experience stays natural. 


We measure pilot workload during flight tasks through subjective self-reports. We choose the NASA-TLX, a multi-dimensional scale designed to obtain workload estimates from operators while they are performing a task or immediately afterward \cite{hart2006nasa}. This scale has been validated and demonstrated to be effective for both fixed-wing and rotocraft pilot workload \cite{battiste1988transport, mansikka2019comparison}. Each TLX survey has 6 questions, asking the participant to rate their mental demand, physical demand, temporal demand, performance, effort, and frustration on a scale from 1-20. We implement a Unity application so that the participants can conveniently input their answers on a Microsoft Surface tablet.

There are 23 flight tasks and the simulation needs to be paused during answering of NASA-TLX. Therefore, performing one NASA-TLX for each individual flight task would likely lead to over-surveying and inaccurate self-reports \cite{cycyota2006not}. This would also break immersion, which could negatively impact the realism of the simulation. To mitigate this, we group consecutive flight tasks with similar expected difficulties together into a single ``task group''. Furthermore, we conduct surveys in batches, with multiple surveys conducted at each pause. The exact way we determine task groups and survey points is shown in Table \ref{tab:flighttasks}. In the end, participants each complete 13 NASA-TLX's over 5 pause points.

\section{MULTIMODAL WORKLOAD ESTIMATION}\label{sec:signal-processing}
Through signal processing and feature extraction, we prepare the collected multimodal signals for statistical analysis and machine learning. Out of the 28 total participants, 15 had complete data with no missing modalities, 2 had inaccurate eye gaze data, 10 had one modality missing, and 1 had two modalities missing. Missing data was attributed mostly to hardware issues, with a few experimenter oversights and software issues. Rather than discarding data from the incomplete participants, we design models that can handle missing values, as will be discussed in Section~\ref{sec:workload}.
\subsection{fNIRS Signal Processing} 
The brain activity signals contain much noise and we follow the standard process \cite{ayaz2018functional} to filter it out. We utilize the fNIR Soft Pro software from Biopac because it has tuned default parameters and convenient functions. We first apply a low-pass filter with a frequency threshold of 2Hz. Then, we apply a motion artifact rejection (SMAR), which is an algorithm that uses sliding window signals from the accelerometer on the fNIRS device to level the optode readings \cite{ayaz2010sliding}. From the refined data, we compute the oxygenation through the Modified Beer Lambert Law (MBLL) \cite{kocsis2006modified}.

\subsection{Eye Movement Processing}
Eye movement can be broken down into saccades and fixations. Saccades are when gaze quickly switches from one target to another, and fixations are when the gaze is stagnant in one place. Previous research \cite{ellis2010eye} has shown that saccadic distance tends to be shorter when pilots perceive their workload as high during flight deck operations, and vice versa. Hence, we hypothesize that saccade and fixation information could be useful features and performed some calculations to extract them.

One way to detect saccades is to look at the speed of gaze position and mark the current timestamp as a saccade if it exceeds a certain threshold \cite{saccade_velocity}. In our processing pipeline, we compute the speed of gaze position at each point in time and determine whether it exceeds an appropriate saccade threshold, determined by manually reviewing footage. The saccade event timestamps are then used to generate three new features: average saccades per second, average fixation time, and average saccade distance.

\subsection{Gaze Annotation}
We hypothesize that the subject of the pilot's gaze could be a useful feature for workload estimation. Thus, we use data collected from the eye tracking glasses, as well as a pre-trained image segmentation model, to determine what the pilot was looking at throughout the experiment.

The Tobii Pro 3 eye-tracking glasses produces a video from the participant's point of view, along with estimated gaze coordinates for each frame. 
We used these measurements to compute the corresponding gaze positions in the screen recording of the flight. This was done using FLANN Based Matching \cite{muja2009fast} with SIFT Descriptors \cite{lowe2004distinctive}.

\begin{figure}[!tbp]
  \centering
  \begin{minipage}[b]{0.49\textwidth}
    \includegraphics[width=\textwidth]{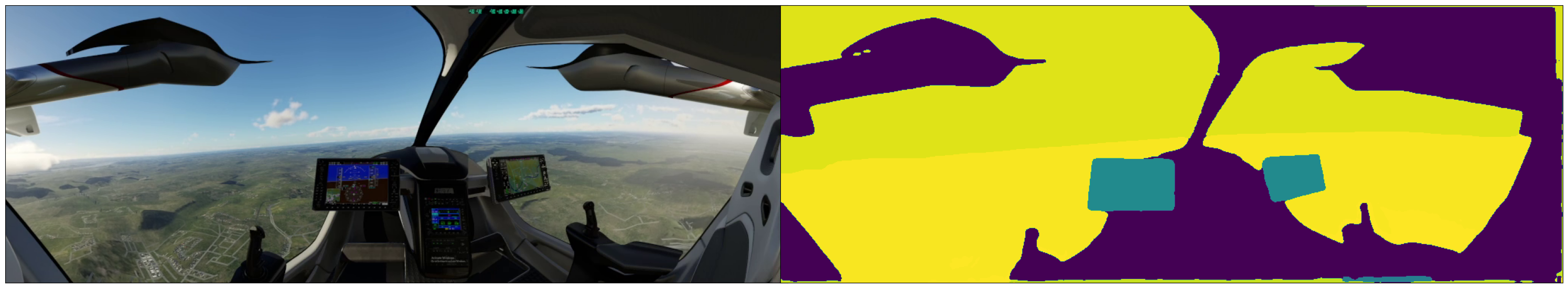}
    \caption{Semantic Segmentation}
    \label{fig:segmentation}
  \end{minipage}\vspace{-0.5cm}
\end{figure}

Next, we use OneFormer \cite{jain2023oneformer}, a transformer-based image segmentation model, to perform semantic segmentation on every frame of our recorded simulation videos from all participants. Specifically, we use OneFormer pre-trained with COCO 2017 \cite{lin2015microsoft} on DiNAT \cite{hassani2023dilated} backbones, and semantic and panoptic classes of COCO 2017. A comparison of the original simulation and the semantic segmentation image is shown in Fig. \ref{fig:segmentation}. To improve accuracy, we override some of segmentations generated by Oneformer using manually segmented static objects that were connected to the plane (and thus didn't move throughout the experiment).

When the pilot’s eye gaze is located at the intersection of multiple objects, it becomes difficult to determine which object the pilot is paying more attention to. Following the method used in \cite{qiu2022incorporating}, we design a semantics priority weighting algorithm. We first group the semantic classes we collected from the segmented frames into eight semantic groups with priority scores: Monitors (10), Roads (7), Buildings (7), Water (5), Sky (4), Ground (4), Airframe (2), and Inside (1). For each eye gaze in the video frame, we extract the semantic groups of 121 pixels in a circular shape from the center of the gaze.

\begin{equation}
    W_{\text{annotation}} = \exp\left(\frac{\text{priority}}{3}\right) - 1 \label{eq:annotation_weight}
\end{equation}

Using Eq. \ref{eq:annotation_weight}, we sum the weights over each class and normalize them to obtain a probability distribution over the 8 semantic classes. This vector is computed for each frame of the original glasses recording.



\subsection{Complete Modality List}
We categorize each data modality into one of four categories: \textbf{Physiological}, \textbf{Behavioral}, \textbf{Situational}, and \textbf{Flight Derivative}, as shown in Table \ref{tab:feature}. Physiological data are biological signals that are correlated with human affective states. Behavioral data deal with the physical behaviors of human subjects. Situational data are the surroundings perceived by the human subjects. Flight Derivative data are quantifiable metrics derived from flight operations, such as velocity and distance.

\begin{table}[!htp]\footnotesize{
  \caption{Feature List}\vspace{-1em}
  \label{tab:feature}
  \centering
  \begin{tabular}{p{1.9cm}p{1.9cm}p{0.5cm}p{3.0cm}}
    \toprule
    Category & Modality & Dim & Comments\\
    \midrule
    \midrule
    Physiological & GSR & 1 & Sweat gland activity\\
    & HR & 2 & HR, IBI\\
    & fNIRS & 16 & Brain hemoglobin levels \\
    & Pupillary & 2 & Pupil diameters \\
    \midrule
    Behavioral & FSR & 2 & Grip force\\
    & Body Pose & 27 & 3D body pose coordinates\\
    & ACC & 3 & Wrist acceleration\\
    & Gaze & 5 & 2D gaze coordinates, saccade, fixation \\
    & Flight Control & 4 & Throttle/joystick inputs\\
    \midrule
    Situational & Gaze Semantics & 8 & Gaze semantic class\\
    \midrule
    Flight Derivative & Flight Derivative & 9 & Aircraft position, orientation, and speed\\
  \bottomrule
\end{tabular}}\vspace{-0.5cm}
\end{table}

\section{STATISTICAL ANALYSIS}\label{sec:statistical-analysis}
We designed and conducted a user study to collect data for workload estimation. All the pilots signed a consent form to participate in the study and were compensated with a \$100 Amazon gift card. The study was approved by the Carnegie Mellon University Institutional Review Board (ID: STUDY2023\_00000131).

\subsection{Demographics} 
\begin{figure}[!htp]
  \centering
  \begin{minipage}[b]{0.45\textwidth}\vspace{-0.2cm}
    \includegraphics[width=\textwidth]{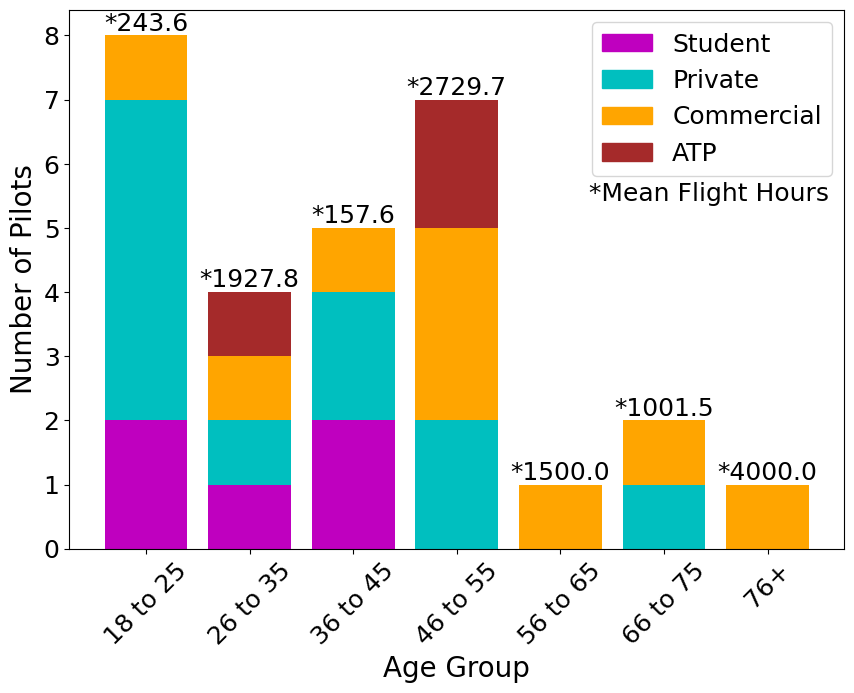}
    \caption{Demographics}
    \label{fig:demographics}
  \end{minipage}\vspace{-1em}
\end{figure}

Figure \ref{fig:demographics} shows a demographic summary of the participants. We recruited a total of 28 pilots based in Pittsburgh, PA. Participants came from a variety of backgrounds with a wide range of ages and flight hours. Based on FAA pilot certification levels, there were 5 student, 11 private, 9 commercial, and 3 airline transport pilots (ATPs). The median age group was 36 to 45, and the average total flight hours was 1,324 with a standard deviation of 2,300. All of the pilots had fixed-wing licenses, and 2 of them also had helicopter experience.

\subsection{NASA-TLX Survey Responses}
\begin{figure}[!htp]
  \centering
  \begin{minipage}[b]{0.49\textwidth}
    \includegraphics[width=\textwidth]{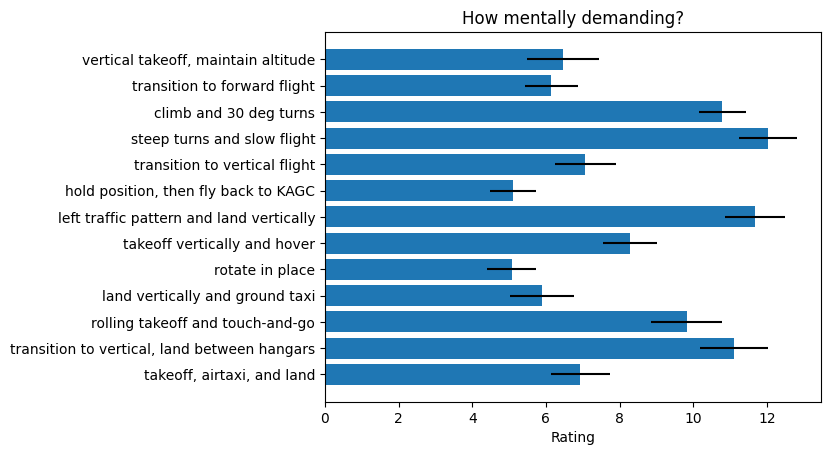}
    \caption{Mean Mental Demand Rating for Each Task Group}
    \label{fig:tlx_dist}
  \end{minipage}\vspace{-0.3em}
\end{figure}

We analyze the distribution of mental demand across all participants for each task. Participants rated mental demand on a scale from 1-20. Figure \ref{fig:tlx_dist} shows the results. On average, the tasks requiring the least mental demand are ``hold position, then fly back to KAGC'' and ``takeoff vertically and hover'', which are rated at around 5. The most difficult tasks are ``steep turns and slow flight'' and ``left traffic pattern and land vertically'', which are rated at around 12. This shows a wide breadth of task difficulty.

As a baseline and sanity check, we attempt to predict participant mental demand using only the pre-screened expected task difficulties. For each participant, we classify their mental demand on each task as low/medium/high. Tasks with mental demand within 0.6 standard deviations from the mean are classified as medium, and the rest are classified as low or high based on which side of the mean they are on. Similarly, we use the pre-screened expected difficulties to categorize tasks as easy if it was 1-2, medium if it was 3-4, and hard if it was 5-6.

\begin{figure}[!tp]
  \centering
  \begin{minipage}[b]{0.45\textwidth}
    \includegraphics[width=\textwidth]{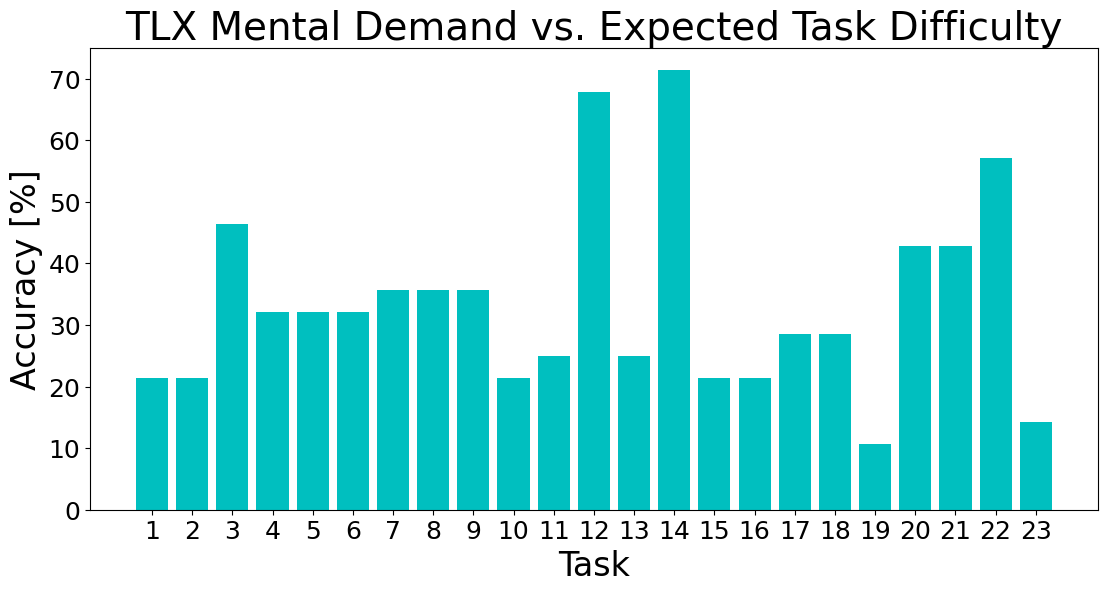}
    \caption{Mental Demand vs. Expected Difficulty}
    \label{fig:mental_demand}
  \end{minipage}\vspace{-1.5em}
\end{figure}

As shown in Fig. \ref{fig:mental_demand}, the assumption that easy tasks result in low mental demand, medium tasks result in medium mental demand, and hard tasks result in high mental demand gives accuracy below 40\% for the most tasks. This suggests that anticipated task difficulty is not a reliable indicator of pilots’ experienced workloads.

\subsection{Exploratory Analysis}
To assess whether the mental demand is related to the physiological or flight technical errors, a linear mixed effects model is used. The analysis allows us to recognize whether psycho-physiological responses impact the pilots' self-reported mental demands. The fixed effects for the model are the features listed in Table \ref{tab:feature}, whereas the random effect is the participant. To shortlist the best fit model, the lowest possible AIC criteria is considered. The analysis is done in R programming language and utilizes the lmer4 package \cite{bates2015package}.

For the physiological responses, significant main effects are found for FSR(thin1) [\emph{F}(1,34.4) = 28.407, p\textless 0.001], standard deviation of left wrist (X) [\emph{F} (1,30.63) = 7.601, p \textless 0.001], standard deviation of left wrist (Y) [\emph{F} (1,32.08) = 7.601, p \textless 
 0.001] and standard deviation of the left wrist (Z) [\emph{F} (1,32.27) = 10.326, p \textless 0.001]. For behavioral responses, the eye glance toward roads is found to have a significant main effect [\emph{F} (1,32.23) = 17.565, p \textless 0.001]. For flight data, main effects are found significant for mean velocity (Z), [\emph{F} (1,89.453) = 6.998, p \textless  0.001], and angular velocity (Y) [\emph{F} (1,95.6)=10.16,p \textless 0.001]. The findings from the analysis confirm that several physiological data modalities and flight behaviors are likely to impact the perceived workload. However, it is critical to estimate workload to understand which set of inputs can provide the appropriate representation of the workload state. The next section describes the machine learning framework for estimating workload models.

\section{WORKLOAD ESTIMATION}\label{sec:workload} 

The input to our machine learning pipeline is aggregated measurements for each modality, and the output is one of three classes: low, medium, or high mental demand. For each modality, we compute the mean, standard deviation, minimum, and maximum of the measured values across each task. To account for individual differences, we also compute normalized versions of the mean, minimum, and maximum based on measurements across the entire experiment. We manually choose a subset of these aggregated measurements to include as input to the models. To handle missing modalities, we use a KNN imputer to fill in any missing values \cite{mahboob2018handling}. This works by finding other data points in the dataset that have similar values for the non-missing modalities, and using those examples to infer the missing ones.

To compute the labels, we first standardize the NASA-TLX mental demand responses across all tasks for each participant. We calculate the mean score and set the low threshold as 0.6 standard deviations below the mean, and the high threshold as 0.6 standard deviations above the mean.

\subsection{Generalized Model}
For the generalized case, our goal is to train a single model that is applicable across all participants. To achieve this, we split our data into training/validation sets across \textit{participants}. In other words, we perform 5-fold cross validation across pilots, resulting in 22-23 pilots in the training set and 5-6 pilots in the validation set for each fold. We average validation accuracy across the 5 folds.
We apply a few popular machine learning algorithms, including linear discriminant analysis (LDA), support vector machine (SVM), random forest (RF), and XGBoost. The balanced accuracies of these models are 42\%, 48\%, 51\%, and 51\%, respectively. The best-performing model is XGBoost, which achieves approximately 51\% accuracy across the 3 classes.

\subsection{Individualized Model}
Although the generalized model outperforms random chance in the 3-class classification, we are able to achieve better results by training models specialized for each participant. In a practical setting, we would have pilots conduct a few test flights to calibrate a model specifically designed to estimate the workload for that particular pilot.

For the individualized models, we choose a single participant as the target and divide their extracted features over \textit{task} boundaries into training and validation sets using 5-fold cross-validation. We upsample the target participant's training set and combine it with a training set derived from every other participant. The purpose of the upsampling is to incentivize the model to specialize on the target pilot by exposing it to a larger volume of their data. Since XGBoost is the best-performing model in the generalized case, we choose to use it for the individualized case as well.

\begin{figure}[!htp]
  \centering
  \begin{minipage}[b]{0.38\textwidth}
    \includegraphics[width=\textwidth]{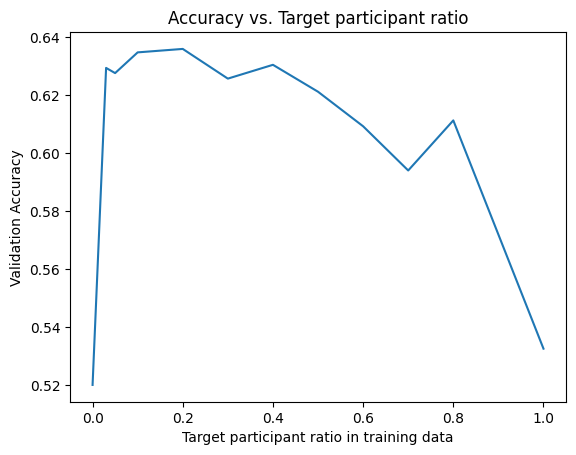}
    \caption{Individualized Model Accuracy}
    \label{fig:XGB_indiv_acc}
  \end{minipage}\vspace{-1.5em}
\end{figure}

Over all possible target participants, we are able to achieve an average of around 63\% balanced accuracy with this method, a boost of 12\% compared to the generalized model. Figure \ref{fig:XGB_indiv_acc} shows how the accuracy changes as we vary the amount of upsampling. When there is no data from the target participant, the individualized model is only able to achieve around 52\% accuracy, which roughly matches the generalized model case. As we increase the proportion of data from the target participant, the model accuracy quickly increases before dropping off. Initially, adding more data from the target participant helps the model specialize on them. However, when too much of the training set is owned by the target participant, the model starts to overfit, which explains the later dropoff in accuracy.

The original dataset consists of 364 input/output pairs over 28 participants, with each participant comprising roughly 3.5\% of the dataset. Upsampling the target participant to 20\% of the dataset yields the best results, increasing the total to 438 input/output pairs.

\subsection{Ablation Study}
\begin{figure}[!htp]
  \centering
  \begin{minipage}[b]{0.42\textwidth}\vspace{-0.2cm}
    \includegraphics[width=\textwidth]{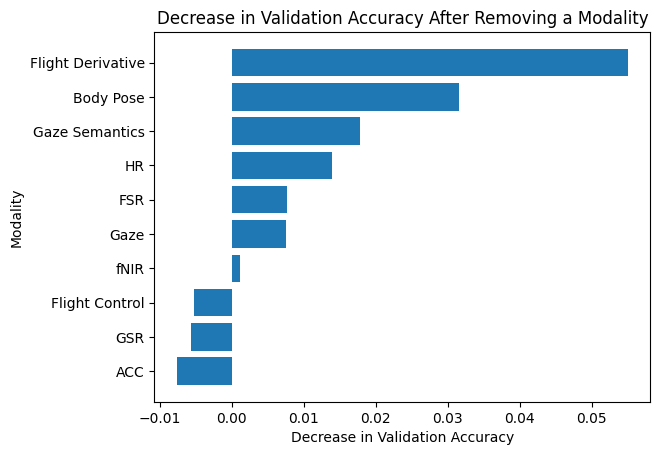}
    \caption{Accuracy Drop by Modality}
    \label{fig:ablation}
  \end{minipage}\vspace{-0.5cm}
\end{figure}

In order to identify which sensors are the most predictive, we perform an ablation study on the sensing modalities to determine which are the best indicators of mental workload. For each modality, we remove it from the inputs and observe the resulting decrease in validation accuracy. As illustrated in Fig. \ref{fig:ablation}, the most important modalities are flight derivative and body pose. Interestingly, flight control, GSR, and ACC have a negative impact on the model's performance.

The ablation study identifies patterns that align with existing research and novel useful sensing modalities for workload estimation.
Flight derivative, naturally directly related to flight maneuver tasks, contains rich data for mental workload. 
GSR, HR, gaze and fNIRS have been repeatably investigated for workload estimation \cite{wilson2002analysis, faulhaber2019crewed, mohanavelu2022machine}. We find HR and gaze to be somewhat predictive, and fNIRS and GSR to be less important.

We are the first in the field to investigate body pose, grip force, and gaze semantics for workload estimation. These sensing modalities all provide surprisingly strong prediction power. Body pose and grip force are related to pilots' flight maneuvers and physical workload \cite{johannes2017psychophysiological}, and thus their correlation with mental workload makes sense. The gaze semantics information conveys which types of objects the pilots were focusing on. This can be explained by a trend we noticed while reviewing experiment footage: pilots tended to stare at the monitors during low workload cruising, but scanned their surroundings rapidly during high-workload landing tasks.

\section{CONCLUSIONS}\label{sec:conclusion}
This work aims to investigate VTOL pilot workload estimation through multimodal sensing. We integrated a simulation platform with XPlane-ROS and implemented a data capture framework including eye gaze, semantics, fNIRS, body pose, grip force, heart rate, and skin conductance signals. We designed a set of flight tasks that represent critical and novel aspects of VTOL operation, such as transition between vertical and horizontal flight, vertical takeoff, and rolling landing. We recruited 28 participants to conduct the experiments and collected the data. Using classical machine learning techniques, we achieved a prediction accuracy of 63\%. This proves the validity of the proposed approach to classify VTOL pilot workload through continuous measurements.

There are some limitations to this research. An ideal goal of this research is to monitor workload continuously, in real time. It is worth noting that we explored a transformer network \cite{wang2023husformer}, with raw time series inputs. However, the transformer model did not out-perform the XGBoost, with a prediction accuracy of around 49\%. 
We attribute the poor performance to limited training data sample size, which led to quick overfitting.
For our future work, we would like to expand the dataset size given more time and resources. More importantly, we would like to explore more sophisticated model structures to potentially achieve time-series raw signal workload predictions. Despite the limitations, this work is the first in the field to estimate VTOL pilot workload. Such a system may provide critical support for VTOL commercial safe operation. The model ablation study also provides insight to VTOL pilot state understanding research, on the most effective sensor sets.




\addtolength{\textheight}{0cm}   




\footnotesize
\bibliographystyle{IEEEbib}
\bibliography{refs}
\balance

\end{document}